\documentclass{PoS}
\usepackage{amsfonts}

\font\twelveeusm=eusm10 scaled 1200
\font\teneusm=eusm10
  \newfam\eusmfam
  \def\eusm{\fam\eusmfam\twelveeusm}
  \textfont\eusmfam=\twelveeusm
  \scriptfont\eusmfam=\teneusm
  \scriptscriptfont\eusmfam=\scriptfont\eusmfam
\font\twelvefrak=eufm10 scaled 1200
\font\tenfrak=eufm10
  \newfam\frakfam
  
  \textfont\frakfam=\twelvefrak
  \scriptfont\frakfam=\tenfrak
  \scriptscriptfont\frakfam=\scriptfont\frakfam
\def\sqr#1#2{{\vcenter{\hrule height.#2pt
   \hbox{\vrule width.#2pt height#1pt \kern#1pt
      \vrule width.#2pt}
   \hrule height.#2pt}}}

\def\bsqr#1#2{{\vrule width #1pt height#2pt}}
\def\bsquare{{\mathchoice\bsqr66\bsqr66\bsqr33\bsqr33}}
\def\badbreak{\penalty1000}

\title{Classifying the Phases of Gauge Theories by Spectral Density of Probing Chiral Quarks}
\ShortTitle{Classifying Gauge Theories by Spectral Density}

\author{Andrei Alexandru\\
        The George Washington University, Washington DC, USA\\
        E-mail:~\email{aalexan@gwu.edu}}
        
\author{\speaker{Ivan Horv\'ath}\\
        University of Kentucky, Lexington, KY, USA\\
        E-mail:~\email{horvath@pa.uky.edu}}

\abstract{
We describe our recent proposal that distinct phases of gauge
theories with fundamental quarks translate into specific types of
low-energy behavior in Dirac spectral density. The resulting scenario
is built around new evidence substantiating the existence of a phase
characterized by bimodal (anomalous) density, and corresponding
to deconfined dynamics with broken valence chiral symmetry. We argue
that such anomalous phase occurs quite generically in these theories,
including in ``real world'' QCD above the crossover temperature, and in
zero-temperature systems with many light flavors.}

\FullConference{The 33rd International Symposium on Lattice Field Theory\\
		14 -18 July 2015\\
		Kobe International Conference Center, Kobe, Japan*}

\begin{document}

\noindent {\bf 1. Introduction.}
The phase structure of gauge theories is actively pursued today,
mainly due to the heavy ion physics programs at RHIC and LHC, and 
as a part of BSM searches related to Higgs discovery. Given its relation 
to strong interactions, SU(3) is particularly interesting in this regard.

Our take on the phases is motivated, in part, by its potential role in
deciphering the vacuum (and thermal state) structure of strong interactions. 
Given that, we work with more generic notion of a phase than what is
typical e.g. in states of matter considerations. Indeed, here it is simply 
a region of parameter space with definite dynamical  property. The phase structure, 
then, is the induced partition of this space via a chosen collection of such 
properties. Notice that, if dynamical features in question are well-defined, 
phase boundaries arise automatically. Transitions mark the corresponding 
changes of dynamics but are not necessarily associated with singularities.

\begin{figure}[t]
\begin{center}
    \centerline{
    \hskip 0.00in
    \includegraphics[width=4.1truecm,angle=0]{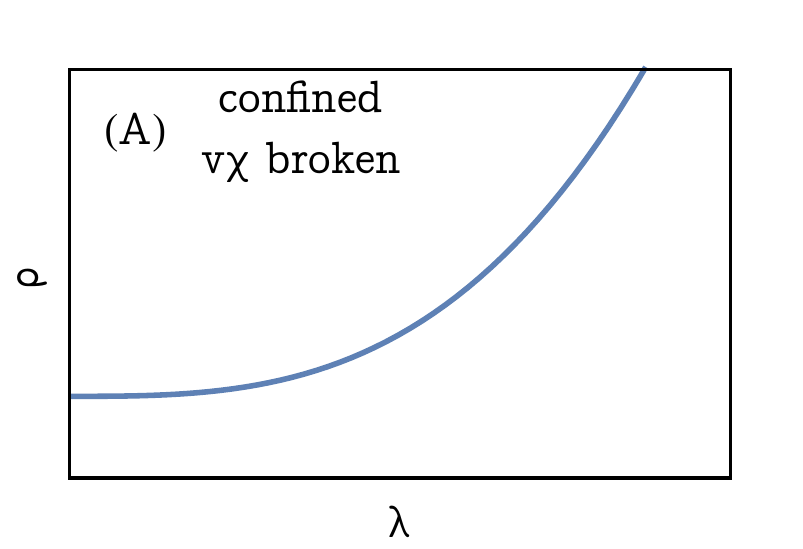}
    \hskip -0.15in
    \includegraphics[width=4.1truecm,angle=0]{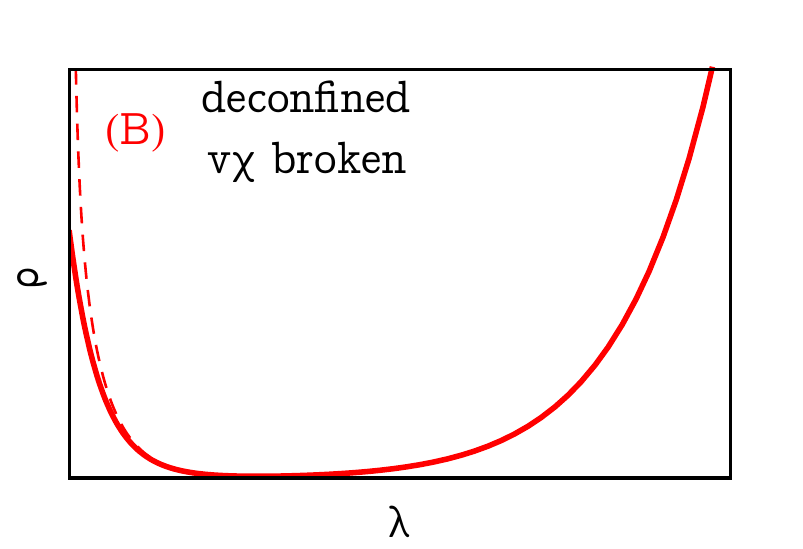}
    \hskip -0.15in
    \includegraphics[width=4.1truecm,angle=0]{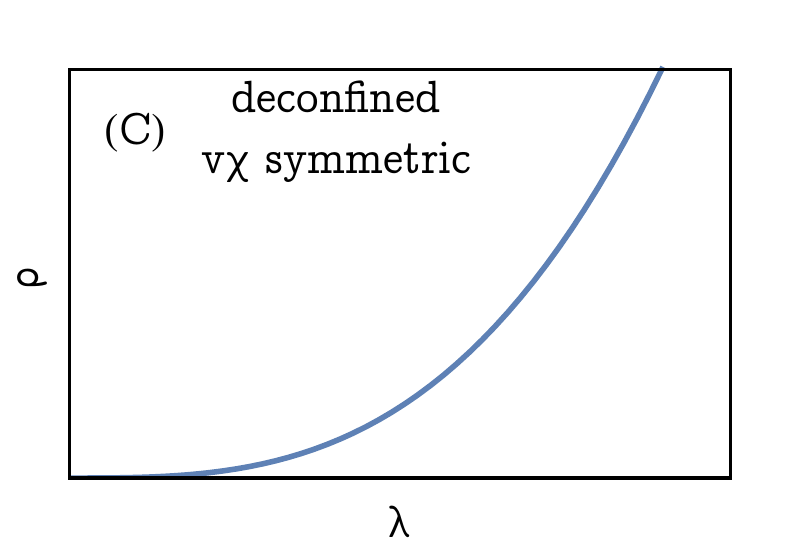}
    \hskip -0.15in
    \includegraphics[width=4.1truecm,angle=0]{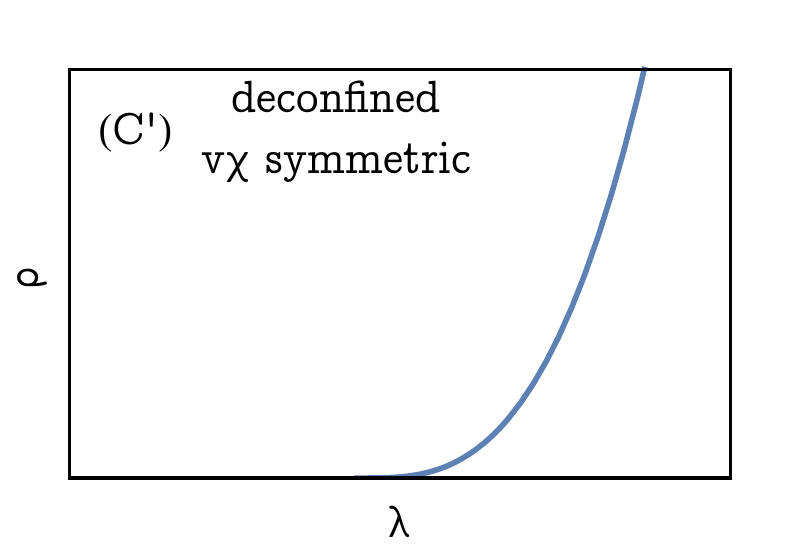}
     }
     \vskip -0.10in
     \caption{Dirac spectral density and confinement/vSChSB structure of ${\eusm T}$.}
     \label{fig:illus1}
    \vskip -0.45in
\end{center}
\end{figure} 

In this talk, the theory space ${\eusm T}$ is spanned by SU(3) 
gauge theories with fundamental quarks: included are cases with arbitrarily 
many flavors of any individual masses, and the system can be thermal.  
We pursue an elementary goal within the above scheme: to design simple 
dynamical properties so that the resulting phase structure relates 
to traditional approach of characterizing vacua in terms of confinement and 
spontaneous chiral symmetry breaking (SChSB) [1].

Our rationale for approaching this is as follows. Consider the gauge 
vacua/thermal states of all theories contained in ${\eusm T}$, and let 
external quarks propagate on the associated backgrounds. If qualitatively 
different dynamical responses are obtained for two different theories, then 
these theories entail qualitatively different vacua or thermal states.

As a general strategy, this is hardly new. Indeed, the classic approach to 
confinement in pure glue case involves {\em infinitely heavy} external 
quark-antiquark pair, and the dependence of their energy on separation. This 
leads to study of extended gauge observables, Wilson loops, and to a well-defined 
criterion for confinement in this special case.

However, as is well known, this criterion fails in theories with dynamical quarks. 
Nevertheless, we intend to keep the idea of probing gauge vacua by external quarks 
even in general case. The only way to proceed then is to let probing quarks have 
finite mass and propagate spatially. In fact, we go to the very opposite mass corner, and 
propose that it is fruitful to probe vacua with quarks that are {\em very light}: always
much lighter than any physical dynamical quarks native to the theory.\footnote{External 
quarks will be frequently referred to as "valence quarks", as is common in lattice 
community.}

What will reflect the qualitative change in dynamics of these light external quarks? 
We suggest that the simplest yet powerful indicator of {\em both} confinement and SChSB 
is the infrared end of Dirac spectral density~\cite{Ale15A}. This crucial element 
of the proposal needs to be elaborated upon first.

\smallskip
\noindent {\bf 2. Dirac Spectral Density and the Claim.}
One can easily see why Dirac spectral density is a reasonable guess if one
tries to characterize dynamical response of external quarks (field $\eta$). After all, 
Dirac operator defines their dynamics via interaction 
$\bar{\eta} (D + m_v) \eta$, where $m_v$ is the mass of a probing quark
as opposed to dynamical quark masses $m_f$.
Moreover, the Dirac matrix naturally involves scale dependence via its 
spectrum $D \psi_\lambda = i \lambda \psi_\lambda$. Thus, to distinguish infrared 
from ultraviolet, we reduce $D$ to the simplest gauge invariant object retaining this 
scale dependence. The result is just the Dirac spectral density $\rho(\lambda)$, namely
\begin{equation}
  \rho(\lambda,m_f,V)  \equiv  \frac{\partial}{\partial \lambda} \, \sigma(\lambda,m_f,V)  
  \qquad\quad
  \sigma(\lambda,m_f,V) \equiv \frac{1}{V} \, 
  \langle \, \sum_{0 \le \lambda_k < \lambda} \,1\;\rangle_{m_f,V}   
  \label{eq:20}
\end{equation} 
Note that, in this context, spectral density is treated as a gauge vacuum object assigned 
to each theory in ${\eusm T}$. The dynamics of very light external quarks is dominated by 
infrared part of the Dirac spectrum, which will provide us with needed dynamical signatures.

Another hint in favor of Dirac spectral density is that mode condensation 
($\rho(\lambda \!\to\! 0)\!>\!0$) indicates valence chiral symmetry breaking 
(vSChSB, $\bar{\eta} \eta \neq 0$). Strictly infrared limit of $\rho(\lambda)$ is thus 
already known to contain part of relevant information: gauge vacuum able to support 
valence condensate is qualitatively different from one that does not. 

\begin{figure}[b]
\begin{center}
    \centerline{
    \hskip 0.00in
    \includegraphics[width=5.6truecm,angle=0]{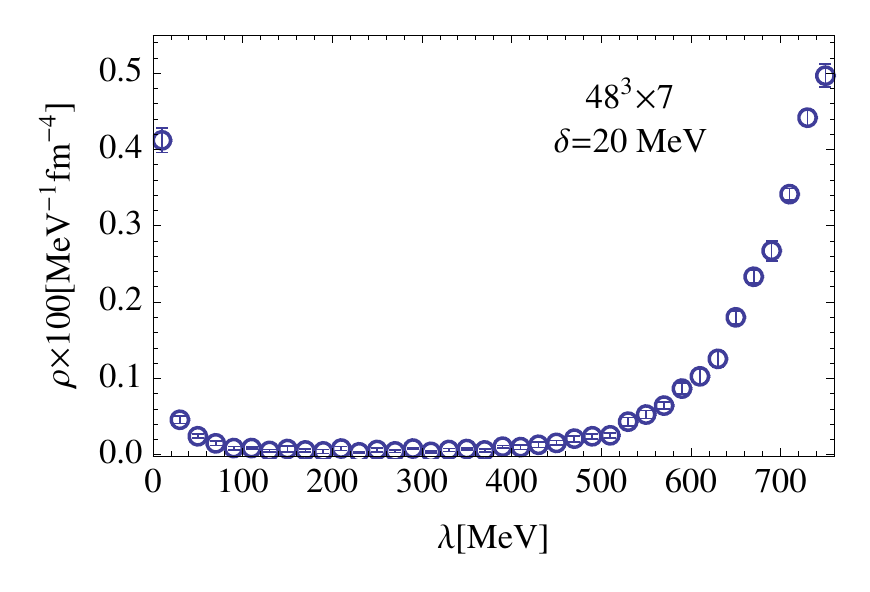}
    \hskip 0.03in
    \includegraphics[width=5.6truecm,angle=0]{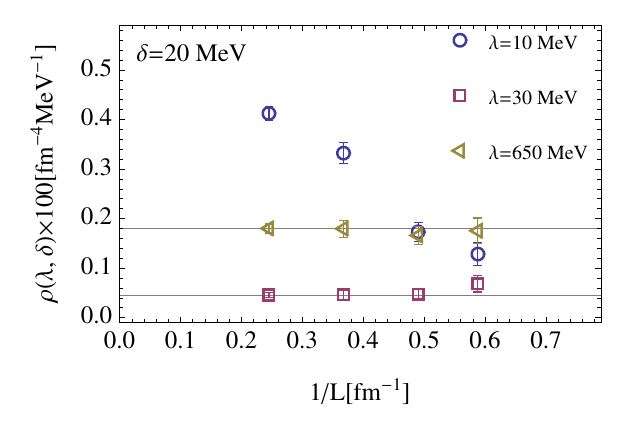}
     }
     \vskip -0.1in
    \centerline{
    \hskip 0.00in
    \includegraphics[width=5.6truecm,angle=0]{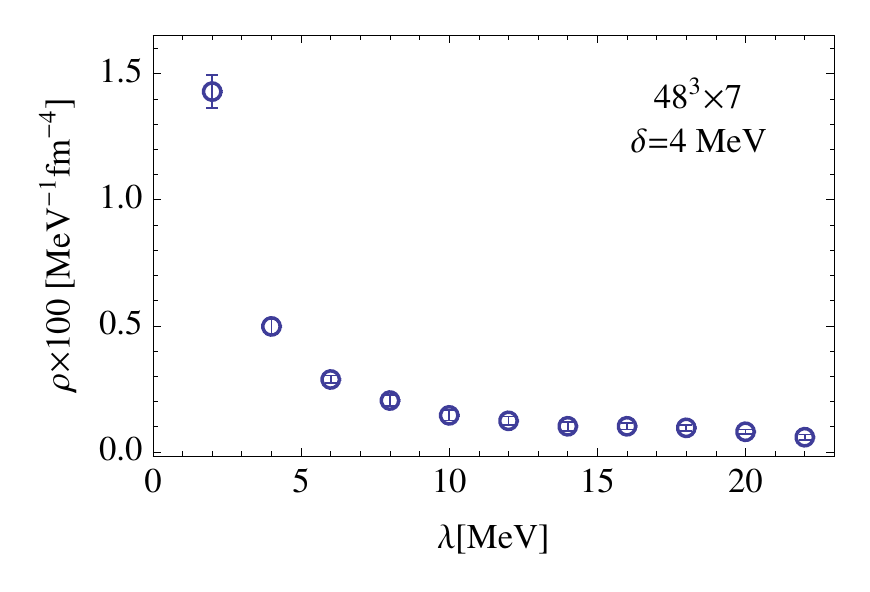}    
    \hskip 0.03in
    \includegraphics[width=5.6truecm,angle=0]{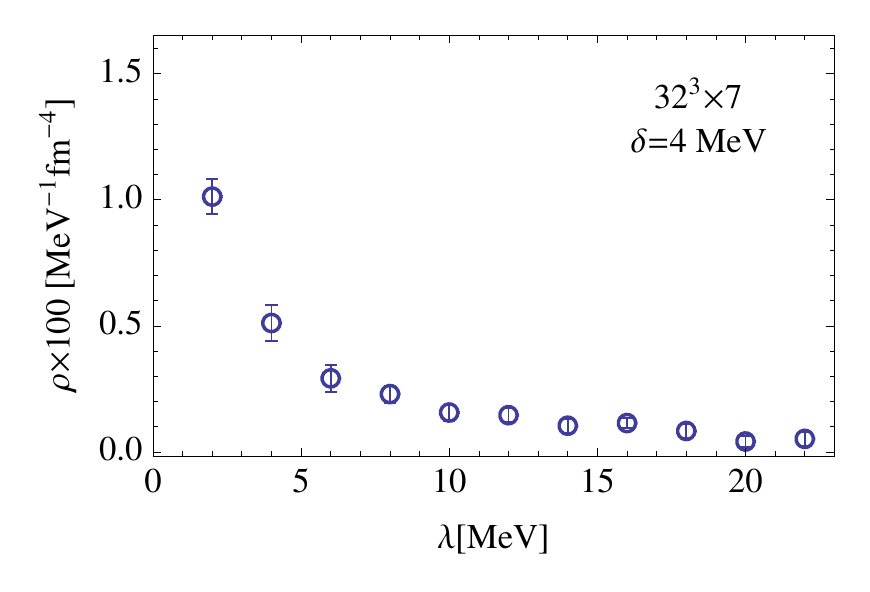}
     }
     \vskip -0.1in
     \caption{Stability of anomalous phase in thermal N$_f$=0 theory under infrared cutoff.}
     \label{fig:infrared}
    \vskip -0.50in
\end{center}
\end{figure} 

Given these considerations, our proposal may be represented by Fig.\ref{fig:illus1}, indicating 
that ${\eusm T}$ is spanned by three behaviors of spectral density~\cite{Ale15A}. The difference 
between the two monotonic types, (A) and (C,C'), is the zero intercept of the latter. The non-standard 
non-monotonic case (B) is referred to as {\em anomalous}, its relevant feature being the {\em bimodal} 
distribution of modes. Existence of such infrared-ulraviolet separation in response of light probing 
quarks to the vacuum of a given theory characterizes this vacuum as being deconfined and (valence) 
chirally broken. Note that there are two types of mode condensation, namely (A) and (B), and thus 
two types of vSChSB: standard and anomalous. However, the proposal entails only one type of 
spectral density reflecting confinement: type (A). This classification implies that phase that is 
simultaneously confined and chirally symmetric doesn't appear. 

Finally, important aspect of this proposal is that the anomalous phase (B) is not a rarity 
in ${\eusm T}$. Rather, we conclude that it occurs generically along paths connecting standard 
phases (A) and (C). This dramatically changes the conventional view of phases in theories 
with fundamental quarks~\cite{Ale15A}.

\smallskip
\noindent {\bf 3. Existence of the Anomalous Phase and Confinement.}
The support of the broad picture described above begins by showing its validity at least
in some part of ${\eusm T}$. This can be done in thermal N$_f$=0 theory where both vSChSB and 
confinement are well-defined. The situation is particularly clean when overlap fermions  
define the dynamics of external quarks, which is what we assume from now on. 
Spectral density of type (B) was first observed in this context some time ago~\cite{Edw99A},
but it was treated as an artifact, its reality not beeing pursued. We thus need to show that 
the phase is stable under both the infrared (volume) and the ultraviolet 
(lattice spacing) cutoffs~\cite{Ale15A}.

To do this, Wilson's lattice theory at T/T$_c$=1.12, set via $r_0 T_c$~\cite{Nec03A},
was used. We reported bimodality of spectral density at this temperature 
previously~\cite{Ale12D,Ale14A}. The ``real Polyakov line" vacuum was always selected in 
spectral calculations with
overlap operator, whose details are described in Ref.~\cite{Ale15A}.  To check infrared 
stability, we simulated $N^3 \!\times\! 7$ systems at $N\!=\!20,24,32,48$ and gauge coupling
corresponding to $a\!=\!0.085 \, \mbox{\rm fm}$ to set the desired temperature. 
Bimodal spectral density, exemplified by top-left plot of Fig.~\ref{fig:infrared}, was obtained 
at all volumes.  Note that $\delta$ is the spectral coarse-graining parameter (bin width).

The density shows very good volume scaling, as seen in the top-right plot. The most 
infrared bin follows a growing trend though, ensuring that the anomalous shape (B) 
results in the infinite-volume limit. The bottom-left plot in Fig.~\ref{fig:infrared} 
shows the closeup on this first bin ($\delta\!=\!4$ MeV), revealing a rapidly decreasing function
whose shape again doesn't  change with infrared cutoff (bottom-right). Indeed, increasing 
the volume only makes the bimodal feature more robust. Our conclusion is that bimodal 
$\rho(\lambda)$ of type (B) is the property of the infinite-volume limit. 

\begin{figure}[b]
\begin{center}
    \centerline{
    \hskip 0.00in
    \includegraphics[width=5.6truecm,angle=0]{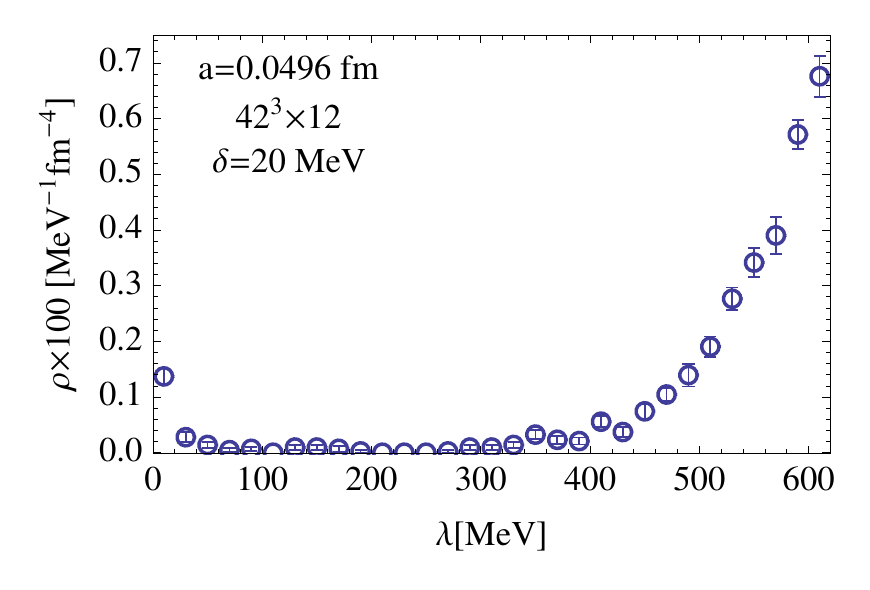}
    \hskip 0.13in
    \includegraphics[width=5.6truecm,angle=0]{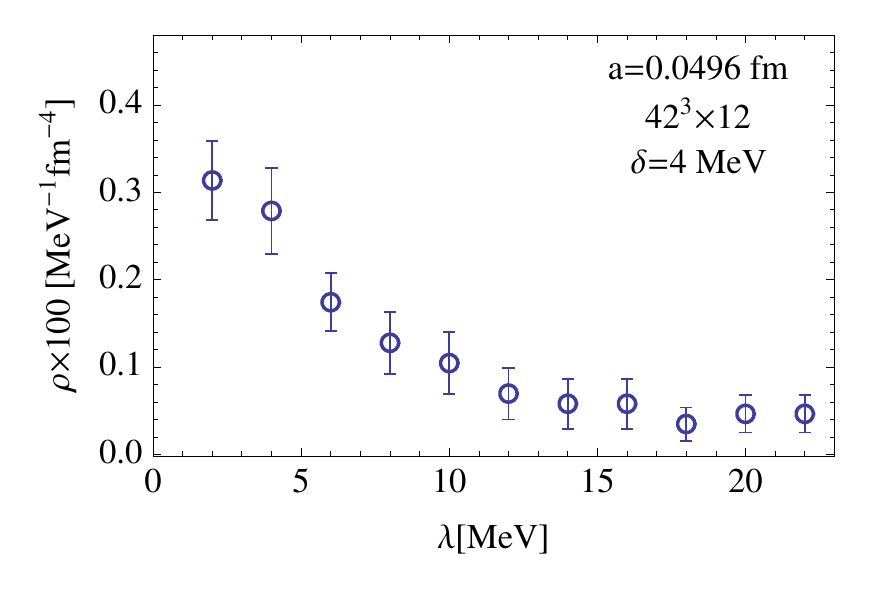}
     }
     \vskip -0.05in
    \centerline{
    \hskip 0.00in
    \raisebox{5pt}{\includegraphics[width=5.7truecm,angle=0]{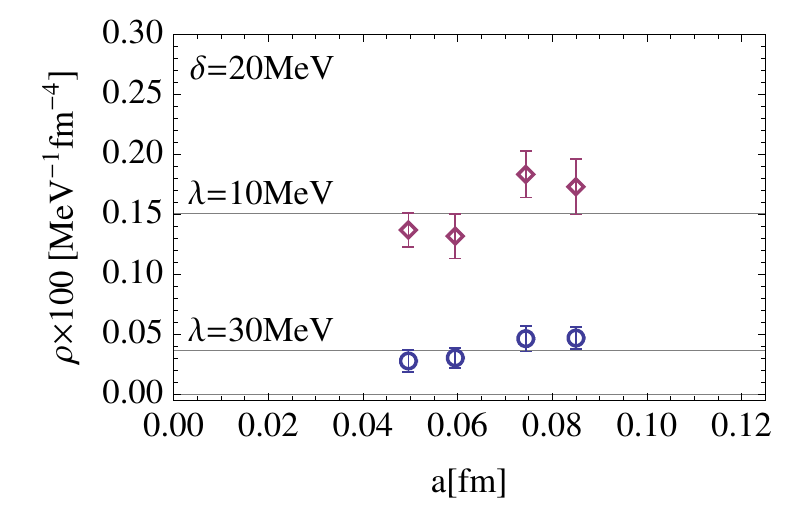}}
    \hskip 0.13in
    \raisebox{4pt}{\includegraphics[width=5.7truecm,angle=0]{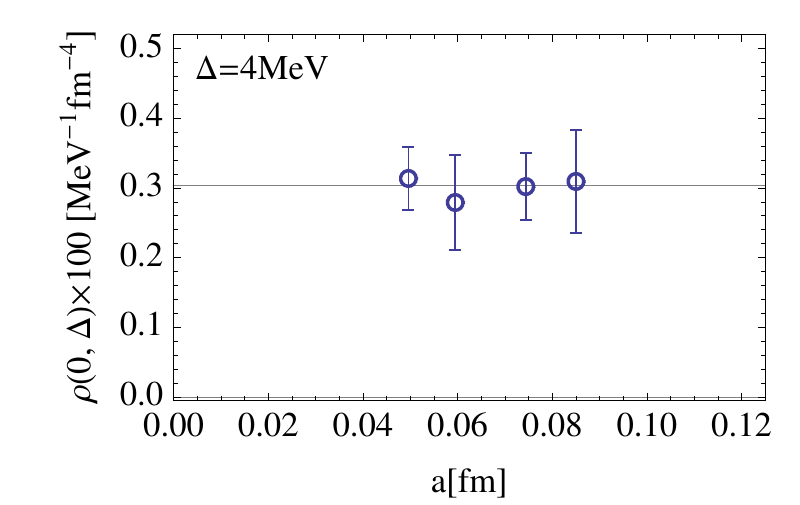}}
     }
     \vskip -0.1in
     \caption{Stability of anomalous phase in thermal N$_f$=0 theory under ultraviolet cutoff.}
     \label{fig:conf_ultraviolet}
    \vskip -0.40in
\end{center}
\end{figure} 

Regarding the continuum limit, we fixed the physical volume to that of $N\!=\!24$ theory,  
and increased ultraviolet cutoff while keeping the temperature fixed. This produced lattices 
$N \!\times\! N_t \!=\! 24 \!\times\! 7,\, 28 \!\times\! 8,\, 34 \!\times\! 10,\, 42 \!\times\! 12$ 
with $a\!=\!0.0850, 0.0744, 0.0595, 0.0496 \,\mbox{\rm fm}$ respectively. Spectral densities were 
found to be bimodal at all cutoffs. The global view and the closeup at finest lattice spacing 
are shown in Fig.~\ref{fig:conf_ultraviolet} (top). Varying ultraviolet cutoff should not significantly 
affect deeply infrared scales unless there is a transition separating lattice and continuum-like 
behaviors. This is not observed as evidenced by lower-left plot showing the scaling of first two bins 
at $\delta\!=\!20 \, \mbox{\rm MeV}$: rather, the results indicate that the bimodal shape will persist 
into the continuum limit. The estimates of mode condensate from very near-zeromodes i.e. 
$\rho(0,\Delta) \!\equiv\! \sigma(0,\Delta)/\Delta$ with $\Delta\!=\!4 \, \mbox{\rm MeV}$, are
stable as well (lower right), confirming the absence of any qualitative change. These results 
substantiate the conclusion that {\em continuum} N$_f$=0 theory at $T/T_c\!=\!1.12$ 
is in the anomalous phase.

\begin{figure}[t]
\begin{center}
    \centerline{
    \hskip 0.00in
    \includegraphics[width=5.8truecm,angle=0]{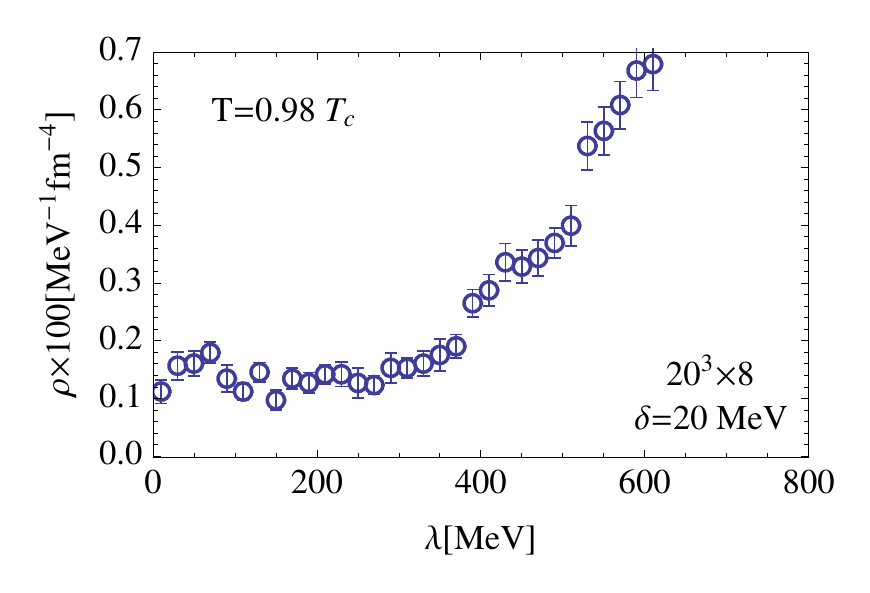}
    \hskip 0.13in
    \includegraphics[width=5.8truecm,angle=0]{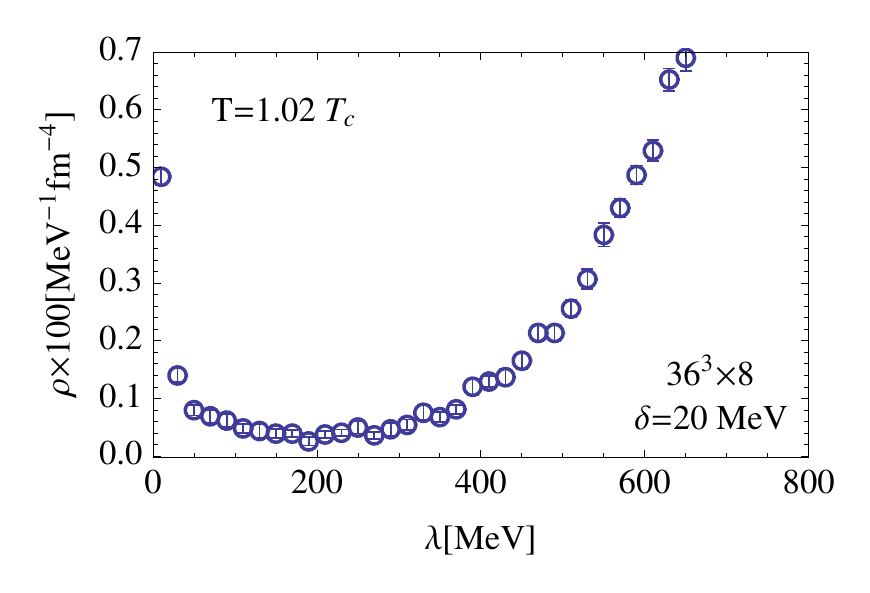}
     }
     \vskip -0.1in
     \caption{Deconfinement and transition to anomalous phase in N$_f$=0 theory.}
     \label{fig:conf_anom}
    \vskip -0.40in
\end{center}
\end{figure}

Finally, it is necessary to demonstrate the connection of anomalous phase to deconfinement, 
in this theory i.e. that the transition from (A) to (B) occurs at deconfinement temperature $T_c$. 
To do that, we simulated the system just below and above $T_c$
($T\!=\!0.98 T_c$ and $T\!=\!1.02 T_c$). The result is shown in Fig.~\ref{fig:conf_anom}, 
indicating that the onsets of deconfinement and anomalous phase indeed coincide.

\smallskip
\noindent {\bf 4. ``Real World'' QCD at Finite Temperature.} 
From the physics standpoint it is clearly important to  establish whether nature's quarks 
and gluons can be in the anomalous phase. The vacuum of N$_f$=2+1 QCD at {\em physical} quark 
masses is a very precise representation of the true strong vacuum. The crossover nature of thermal 
transition in strong interactions was in fact worked out within this framework~\cite{Aok06A}.
To see whether anomalous phase is encountered in the crossover region, we calculated
overlap Dirac spectra in the same setting.

\begin{figure}[b]
\begin{center}
    \centerline{
    \hskip 0.00in
    \includegraphics[width=5.8truecm,angle=0]{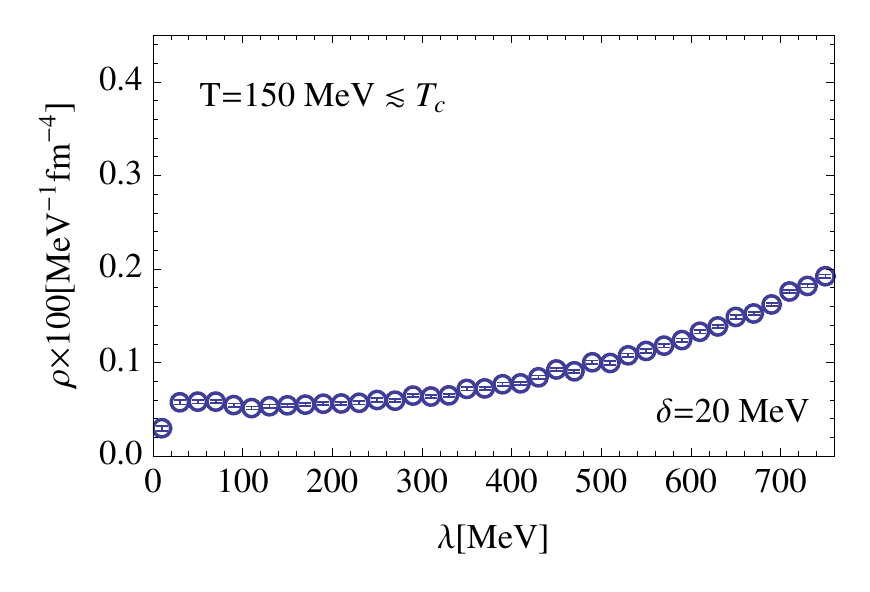}
    \hskip -0.13in
    \includegraphics[width=5.8truecm,angle=0]{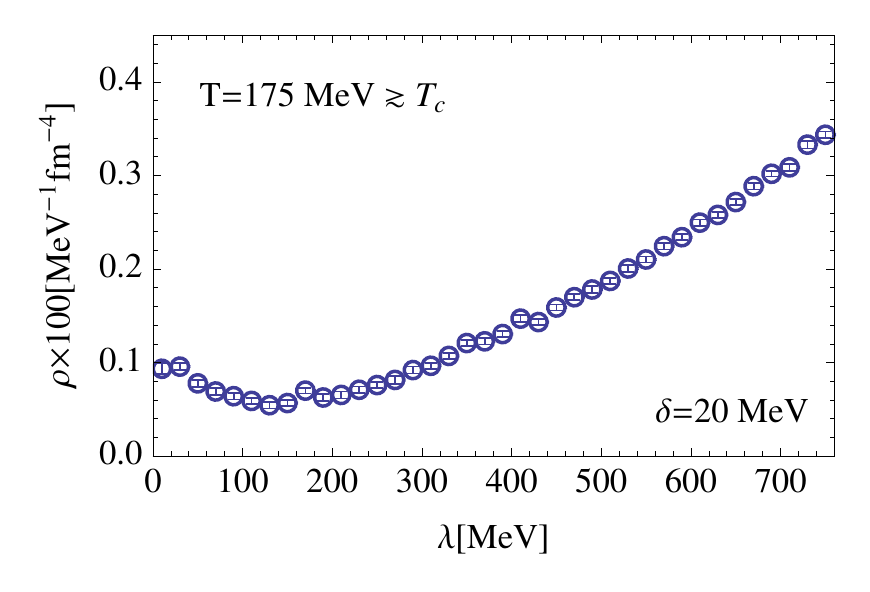}
    \hskip -0.13in
    \includegraphics[width=5.8truecm,angle=0]{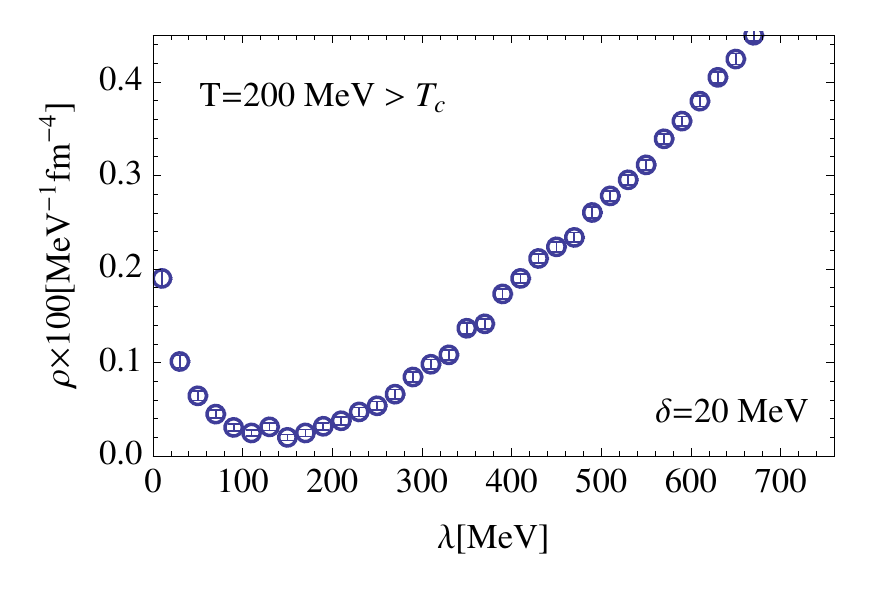}
     }
     \vskip -0.1in
     \caption{Anomalous phase across the thermal crossover in N$_f$=2+1 QCD at physical 
     point.}
     \label{fig:nf2p1}
    \vskip -0.40in
\end{center}
\end{figure}

Computations were performed on $32^3\times 8$ lattices of stout-improved staggered 
fermions and Symanzik-improved glue (see Ref.~\cite{Bor10A} for details). 
Transition temperature is not a unique concept at crossover, and depends on the defining 
observable. In this case, the temperature from $\bar{\psi} \psi$ was reported to be 
$T_c \approx 155 \, \mbox{\rm MeV}$, and from Polyakov line 
$T_c \approx 170 \, \mbox{\rm MeV}$. We thus chose ensembles at temperatures 
$T\!=\!150,\,175$ and 200 MeV, covering the crossover region and beyond.

Computed spectral densities are shown in Fig.~\ref{fig:nf2p1}. One can see that just below 
the lower edge of the crossover (left), $\rho(\lambda)$ exhibits the expected (A) behavior, 
while at the upper edge (middle), the anomalous phase is present.\footnote{Thermal anomalous 
phase in similar setting was also seen in Ref.~\cite{Dic15A}.} Heating the system 
well past the crossover region ($T\!=\!200 \, \mbox{\rm MeV}$), the bimodal nature of 
the spectrum completely takes over (right). Since the ensembles used were extensively 
checked for lattice artifacts, the above evidence leads us to conclude that the anomalous 
phase $T_c \!<\! T \!<\! T_{ch}$, is a feature of strong interactions, and the proposed general 
scheme is obeyed. Here $T_c$ is defined by the onset of anomalous phase, and 
$T_{ch}$ marks the valence chiral restoration temperature.

\smallskip
\noindent {\bf 5. Anomalous Phase via Light Quarks.}
Somewhat surprisingly, it was recently found that the effects of light dynamical quarks alone,
without thermal bath,  generate anomalous phases at sufficiently large number of 
flavors~\cite{Ale14A,Ale14C}. Specifically, in N$_f$=12 theory with staggered fermions, it was 
observed that there is a mass $m_c$ below which the anomalous phase appears. 
The bimodality of Dirac spectrum either extends to arbitrary small masses, or there is  
a non-zero $m_{ch}\!<\!m_c$ below which valence chiral symmetry gets restored. In this way, 
mass effects are analogous to those of temperature.

\begin{figure}[t]
\begin{center}
    \centerline{
    \hskip 0.00in
    \includegraphics[width=5.8truecm,angle=0]{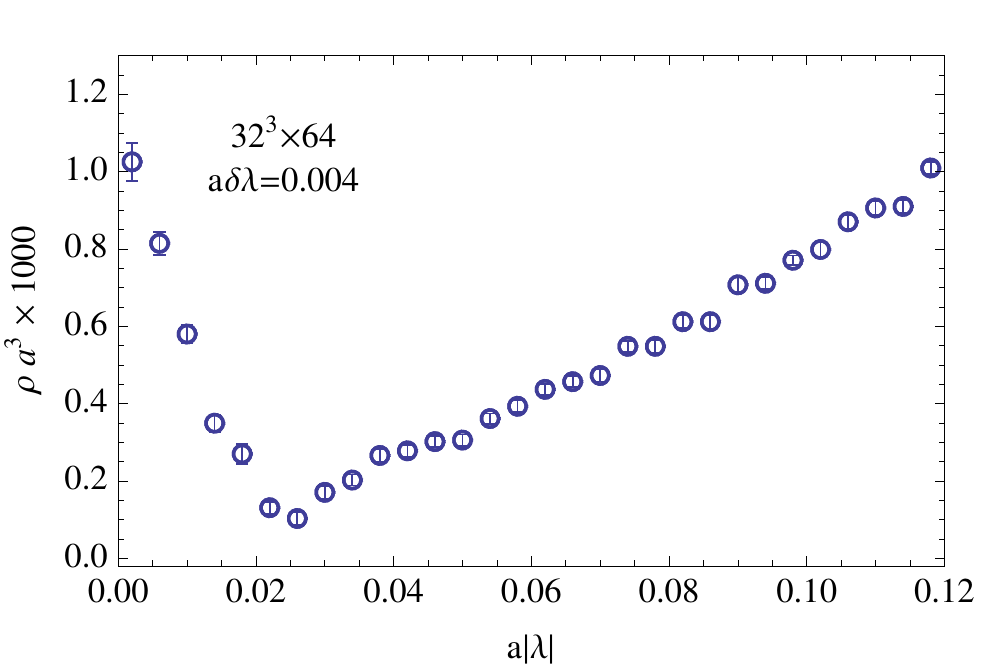}
    \hskip 0.13in
    \includegraphics[width=5.8truecm,angle=0]{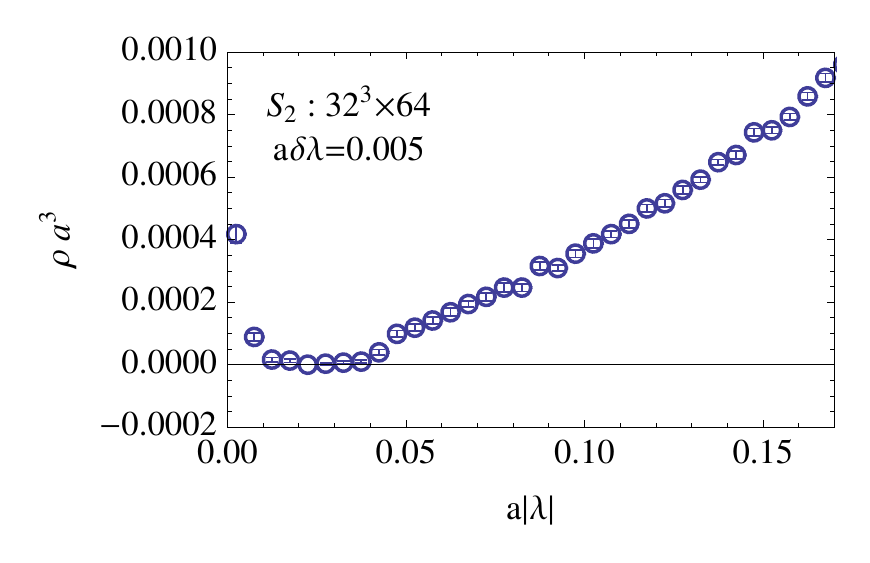}
     }
     \vskip -0.1in
     \caption{Anomalous spectral densities in N$_f$=8 (left) and N$_f$=12 
     (right)~\cite{Ale14A} theories.}
     \label{fig:light_flavs}
    \vskip -0.40in
\end{center}
\end{figure}

It is interesting to determine the minimal number of flavors realizing the above scenario. To get 
the first insight, we made a preliminarily inquiry into the N$_f$=8 theory.  As in the N$_f$=12 case, 
we deal with staggered nHYP dynamical fermions and the fundamental-adjoint combination in 
gauge action to avoid known lattice artifacts ($\beta_F\!=\!4.8$, $\beta_A/\beta_F\!=\!-0.25$).  
The simulation was carried out at zero mass, with finite volume providing for a small effective gap 
in the staggered spectrum. The result is shown in Fig.~\ref{fig:light_flavs} (left). 
Clearly, strong bimodality is observed in this case as well. The anomalous spectrum for N$_f$=12 
($am\!=\!0.0025$, $\beta_F\!=\!2.4$) is also shown~\cite{Ale14A}, although a well-defined 
comparison (equivalent volumes, couplings and effective masses) is obviously difficult to arrange.

\smallskip
\noindent {\bf 6. The Generalization.}
We showed that both light-quark effects (low masses, large N$_f$) and thermal 
effects (large $T$) lead to type (B) anomalous phases. However, these are the only 
available tuning knobs to change the nature of vacuum/thermal state in ${\eusm T}$. We 
thus propose that anomalous phases occur generically on paths connecting chirally broken 
and chirally symmetric vacua. Moreover, the presented evidence for coincidence of anomalous 
and deconfined/chirally broken phases in relevant parts of ${\eusm T}$ leads us to 
extend the correspondence as described in Sec.~2. 

\begin{figure}[t]
\begin{center}
    \centerline{
    \hskip 0.00in
    \includegraphics[width=6.8truecm,angle=0]{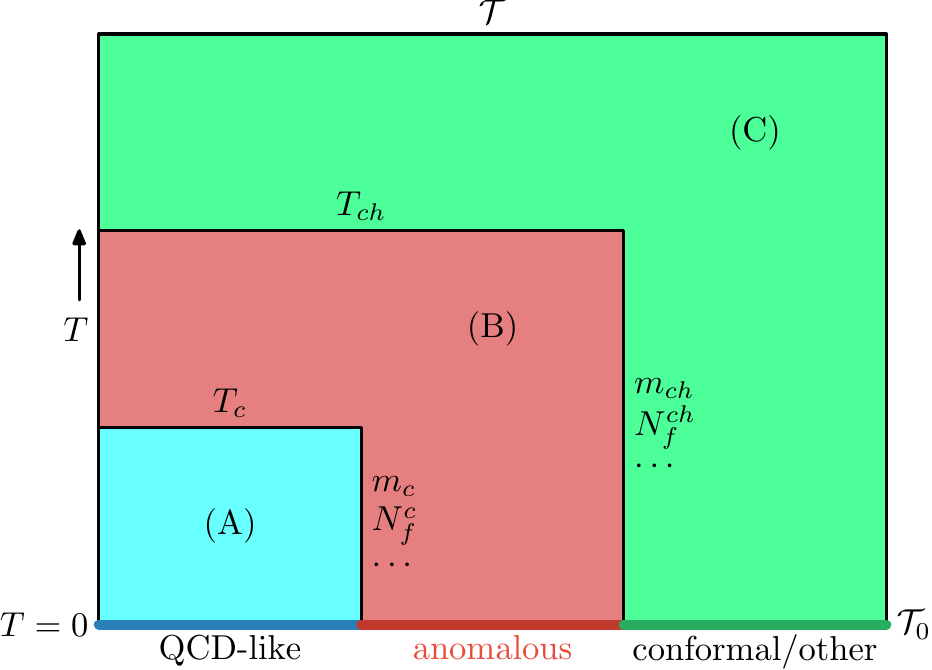}
    \hskip 0.50in
    \includegraphics[width=6.8truecm,angle=0]{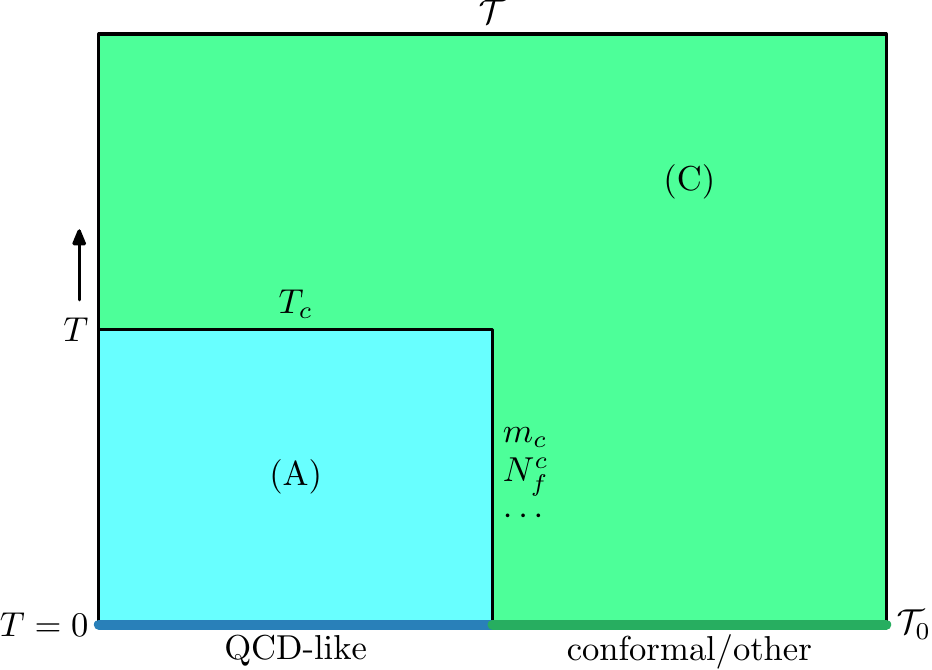}
     }
     \vskip -0.0in
     \caption{The proposed structure of set ${\eusm T}$ (left) compared to the conventional one (right).}
     \label{fig:Tsets}
    \vskip -0.35in
\end{center}
\end{figure}

The proposed scenario brings in a significant change into the standard picture of ${\eusm T}$
in terms of chiral symmetry breaking and confinement. Indeed, the contrast can be 
represented by Fig.~\ref{fig:Tsets}. In the new scenario (left) the anomalous phase
is wedged between the standard confined/broken and deconfined/symmetric phases (A) and (C).
With other parameters fixed, the anomalous regimes $T_c \!<\! T\! <\! T_{ch}$ and
$m_{ch} \!<\! m \!<\! m_c$ are particular cases of this generic behavior. 

The special case we wish to explicitly highlight in this regard is that of $N_f$ massless flavors 
at zero temperature. Connecting the $N_f\!=\!2$ theory believed to be confined, chirally broken 
and type (A), to a theory at sufficiently high $N_f > N_f^{cr}$ so that it is deconfined, chirally
symmetric and type (C), we raise the possibility of anomalous phase~\cite{Ale15A}   
\begin{equation}
    2 < \mbox{\rm N}_f^c < \mbox{\rm N}_f < \mbox{\rm N}_f^{ch} = 
    \mbox{\rm N}_f^{cr}
\end{equation}
Here $N_f^{cr}$ is the usual lower edge for Banks-Zaks infrared fixed point. Thus, in our 
scenario, this "anomalous window" precedes the conformal window. Given the results presented 
here, both $N_f\!=\!8$ and $N_f\!=\!12$ theories may in fact turn out to reside in the former.

\bigskip

\noindent{\bf Acknowledgments:} 
We are indebted to Szabolcs Bors\'anyi, Zolt\'an Fodor, Anna Hasenfratz, David Schaich and 
Mingyang Sun for various help.  A.A. was supported by U.S. National Science Foundation under CAREER 
grant PHY-1151648. I.H. was supported in part by Department of Anesthesiology 
at the University of Kentucky. 

\end{document}